\documentclass[a4paper,11pt]{article}
\usepackage{pos}

\title{Latest Analysis Results from the KASCADE-Grande Data}
\ShortTitle{Results from the KASCADE-Grande}

\author{
D.~ Kang$^{a,*}$,
J.C.~ Arteaga-Vel\'azquez$^{b}$,
M.~ Bertaina$^{c}$,
A.~ Chiavassa$^{c}$,
K.~ Daumiller$^{a}$,
V.~ de Souza$^{d}$,
R.~ Engel$^{a,e}$,
A.~ Gherghel-Lascu$^{f}$,
C.~ Grupen$^{g}$,
A.~ Haungs$^{a}$,
J.R.~ H\"orandel$^{h}$,
T.~ Huege$^{a}$,
K.-H.~ Kampert$^{i}$,
K.~ Link$^{a}$,
H.J.~ Mathes$^{a}$,
S.~ Ostapchenko$^{j}$,
T.~ Pierog$^{a}$,
D.~ Rivera-Rangel$^{b}$,
M.~ Roth$^{a}$,
H.~ Schieler$^{a}$,
F.G.~ Schr\"oder$^{a}$,
O.~ Sima$^{k}$,
A.~ Weindl$^{a}$,
J.~ Wochele$^{a}$,
J.~ Zabierowski$^{l}$
}
\affiliation[a]{Karlsruhe Institute of Technology, Institute for Astroparticle Physics, Germany}
\affiliation[b]{Universidad Michoacana, Instituto de F\'{i}sica y Matem\'{a}ticas, Morelia, Mexico}
\affiliation[c]{Dipartimento di Fisica, Universit\`{a} degli Studi di Torino, Italy}
\affiliation[d]{Universidade S\~{a}o Paulo, Instituto de F\'{i}sica de S\~{a}o Carlos, Brasil}
\affiliation[e]{Karlsruhe Institute of Technology, Institute of Experimental Particle Physics, Germany}
\affiliation[f]{Horia Hulubei National Institute of Physics and Nuclear Engineering, Bucharest, Romania}
\affiliation[g]{Department of Physics, Siegen University, Germany}
\affiliation[h]{Dept. of Astrophysics, Radboud University Nijmegen, The Netherlands}
\affiliation[i]{Fachbereich Physik, Universit\"at Wuppertal, Germany}
\affiliation[j]{Hamburg University, II Institute for Theoretical Physics, 22761 Hamburg}
\affiliation[k]{Department of Physics, University of Bucharest, Bucharest, Romania}
\affiliation[l]{National Centre for Nuclear Research, Department of Astrophysics, Lodz, Poland}
\affiliation[*]{\rm Presenter}

\emailAdd{donghwa.kang@kit.edu}

\abstract{
KASCADE-Grande, the extension of the multi-detector setup of KASCADE, was devoted to measure the properties of extensive air showers initiated by high-energy cosmic rays in the primary energy range of 1~PeV up to 1~EeV. The observations of the energy spectrum and mass composition of cosmic rays contribute with great detail to the understanding of the transition from galactic to extragalactic origin of cosmic rays, and furthermore to validate the properties of hadronic interaction models in the air shower development. Although the experiment is fully dismantled, the analysis of the entire KASCADE-Grande data set continues. We have recently investigated the impact of different post-LHC hadronic interaction models, QGSJETII-04, EPOS-LHC, Sibyll 2.3d, on air shower predictions in terms of the reconstructed spectra of heavy and light primary masses, including systematic uncertainties. In addition, the conversely discussed evolution of the muon content of high-energy air showers in the atmosphere is compared with the predictions of different interaction models. In this contribution, the latest results from the KASCADE-Grande measurements will be discussed.}

\ConferenceLogo{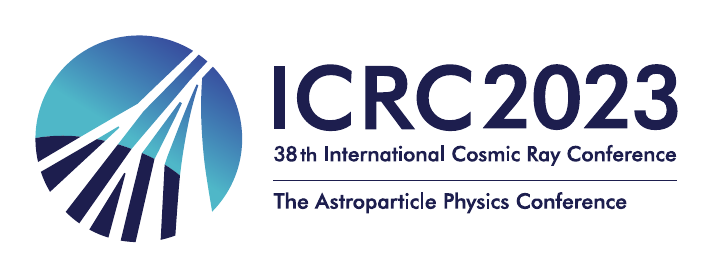}

\FullConference{
  38th International Cosmic Ray Conference (ICRC2023)\\
  26 July - 3 August, 2023\\
  Nagoya, Japan}

\begin{document}
\maketitle

\section{Introduction}
\label{sec:intro}
The KASCADE \cite{KASCADE} and its extension KASCADE-Grande \cite{KASCADE-Grande} experiments were dedicated to understand the chemical mass composition, energy spectrum and arrival direction of high-energy cosmic rays. These measurements are, in particular, contributed to identifying the transition region of galactic and extragalactic cosmic rays.
The experiments were located at the Karlsruhe Institute of Technology, Karlsruhe, Germany (49.1$^{\circ}$ north, 8.4$^{\circ}$ east, 110 m a.s.l.) and measured extensive air showers in the energy range of PeV to EeV. One remark is that a multi-detector setup of KASCADE allowed us to reconstruct the number of electrons and muons, separately, for individual air showers.
Detailed analysis of more than 20 years of measured data presents fruitful results: The reconstructed all-particle energy spectrum measured by KASCADE shows a knee-like structure, mainly, due to a steepening of spectra of light (H and He) elements \cite{KA_APP}.
In respect of the KASCADE-Grande observation, the all-particle energy spectrum \cite{KG_APP} reveals a concave behavior just above 10$^{16}$ eV and a break at around 10$^{17}$ eV, where a knee-like feature would be expected as the knee of the heavy primaries, mainly, iron, in accordance with the rigidity dependence. The knee-like structure in the spectrum of the heavy-mass group is observed significantly at around 80 PeV \cite{KG_PRL}. Furthermore, an ankle-like structure is observed at an energy of 100 PeV in the energy spectrum of light primary cosmic rays \cite{KG_PRD}.

The data collection was completed at the end of 2013, though analysis of more than 20 years measured data is still performed. In particular, we use the data to investigate the validity of new and updated hadronic interaction models. Moreover, the KASCADE and KASCADE-Grande experiments were able to measure the number of muons with high accuracy. Measurements of KASCADE-Grande on the muon number suggest that the attenuation length of muons in the atmosphere is larger than the predictions from the hadronic interaction models. In order to investigate the muon anomalies, we also estimated the number of muons as a function of the primary energy at different zenith angles using the data from KASCADE-Grande.

In this paper, we will discuss on recently ongoing studies related to the updated energy spectra of heavy and light mass composition groups including systematic uncertainties and the study of the muon content \cite{JuanCarlos}. 

\section{Spectra of heavy and light mass groups}
\subsection{Data sets}
The analysis presented here is only based on the KASCADE-Grande data. After applying quality cuts, a total of $1.7\times10^{7}$ events for a measuring time of 1865.62 days were used in this analysis. The zenith angle is used up to 40$^{\circ}$.
Full trigger and reconstruction efﬁciency for the shower size ($N_{ch}$) is reached at the number of charged particles of around $10^{6}$, which corresponds to a primary energy of about $10^{16}$~eV.
For the air shower simulations, the CORSIKA \cite{CORSIKA} program has been used, applying different embedded hadronic interaction models. The FLUKA (E < 200 GeV) model has been used for hadronic interactions at low energies, while high-energy interactions were treated with QGSJET-II-04 \cite{QGS04}, EPOS-LHC\cite{EPOS} and SIBYLL 2.3d \cite{SIB23d}. Air showers induced by five different primaries (H, He, C, Si, and Fe nuclei) have been simulated. The simulations cover the energy range of 10$^{14}$ - 10$^{18}$~eV
with zenith angles in the interval 0$^{\circ}$- 42$^{\circ}$. The spectral index in the simulations was -2 and for the analysis it is weighted to a slope of -3.

\subsection{Separation into mass groups}
To reconstruct energy spectra for individual mass groups, the full dataset is divided into two subsets of heavy and light primary groups, according to the $y_{CIC}$ cut method \cite{KG_APP}. This cut is based on the shower size ratio of the attenuation-corrected muon and charged particle numbers:
$y_{CIC}$ = log$_{10}(N_{\mu, CF})^{CIC}$ / log$_{10}(N_{ch})^{CIC}$,
where the total muon number is corrected for systematic biases. The Constant Intensity Cut (CIC) technique is applied to correct the number of muon and the number of charged particles for the attenuation effect in the atmosphere.
The events satisfying the condition ($y_{CIC} \geq y_{CIC}^{thr}$) are defined as heavy (electron-poor) events and the remaining as light (electron-rich) events. The value of the selection criteria of $y_{CIC}^{thr}$ depends on the interaction models and it is defined to be between the CNO and the silicon elements for each models. Therefore, the heavy component consists mainly of silicon and iron primaries, whereas the light mass groups are proton, helium and carbon nuclei.
Figure~1 shows the $y_{CIC}$ as a function of the primary energy for the hadronic interaction models QGSJET-II-04, EPOS-LHC, and SIBYLL 2.3d. The dashed line in Fig.~1 is the separation criteria for heavy and light mass groups.

\begin{figure}[t]
\centering
\includegraphics[width=0.329\textwidth]{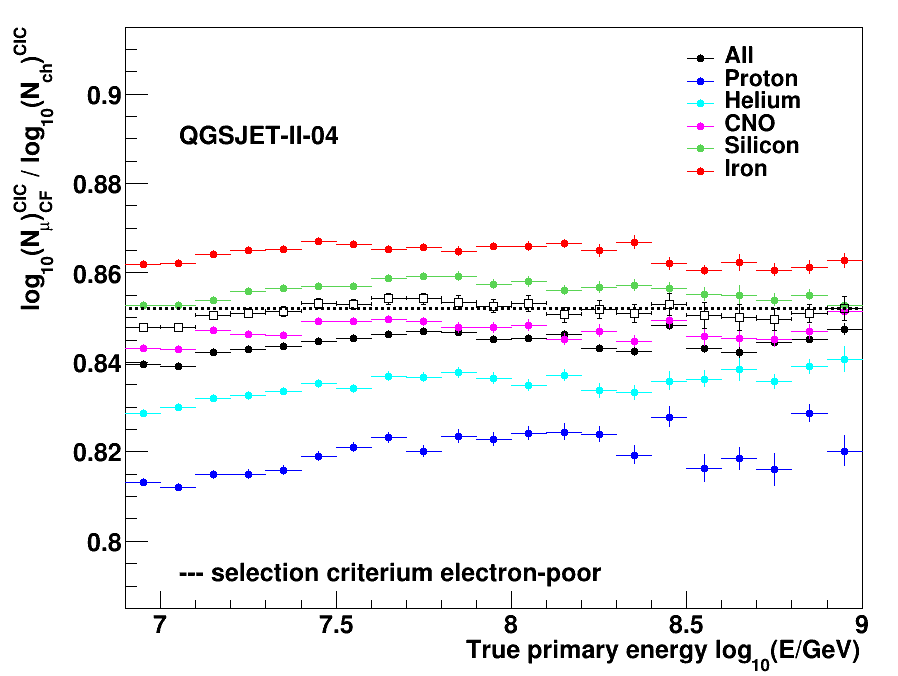}
\includegraphics[width=0.329\textwidth]{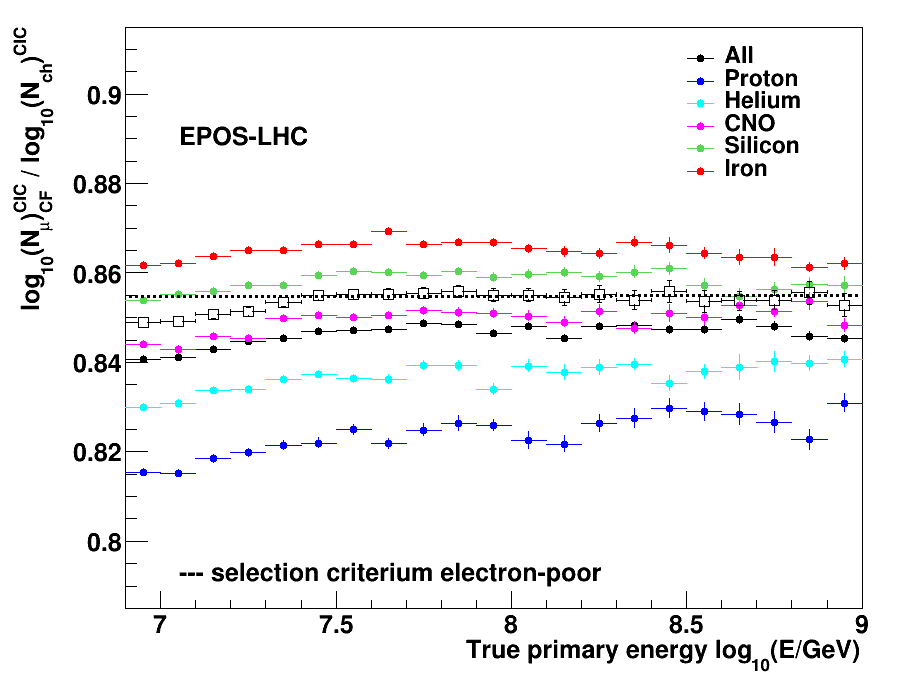}
\includegraphics[width=0.329\textwidth]{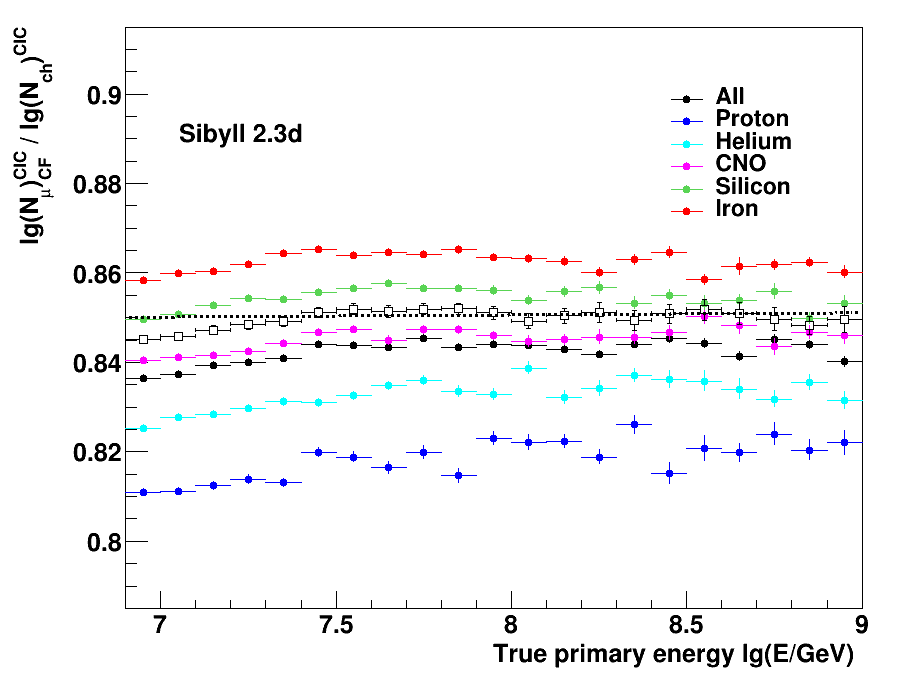}
\caption{
  The $y_{CIC}$ as a function of the primary energy for QGSJET-II-04 (left), EPOS-LHC (middle), and SIBYLL 2.3d (right).
  The dashed line represents the selection criteria for heavy and light mass groups.
  The open squares are the average values of the silicon and CNO mass components.
}
\end{figure}

\begin{table}[b!]
  \begin{center}
   \small 
\begin{tabular}{lccccc}
\hline
Models	      &\multicolumn{2}{c}{Heavy}              & \multicolumn{2}{c}{Light}\\
              & $a$               & $b$               & $a$               & $b$  \\ \hline
QGSJET-II-04  & 0.885 $\pm$ 0.003 & 1.863 $\pm$ 0.023 & 0.936 $\pm$ 0.006 & 1.285 $\pm$ 0.039\\
EPOS-LHC      & 0.885 $\pm$ 0.004 & 1.847 $\pm$ 0.023 & 0.919 $\pm$ 0.005 & 1.388 $\pm$ 0.035\\
SIBYLL 2.3d   & 0.892 $\pm$ 0.004 & 1.802 $\pm$ 0.024 & 0.943 $\pm$ 0.005 & 1.216 $\pm$ 0.035\\ \hline
\end{tabular}
\caption{
  The coefficients of the energy calibration for QGSJET-II-04, EPOS-LHC, and SIBYLL 2.3d
  for heavy and light primaries, respectively.} 
\end{center}
\end{table}

\subsection{Energy calibration}
Figure~2 presents the energy calibration function for light and heavy induced showers. With the assumption of a linear dependence in logarithmic scale (log$_{10}(E/{\rm GeV})$ = $a \cdot$log$_{10}(N_{ch}) + b$), a linear fit is applied in the range of full trigger and reconstruction efficiencies. The resulting coefficients of the energy calibration for QGSJET-II-04, EPOS-LHC and SIBYLL 2.3d are summarized in Table~1.

\begin{figure}[t]
\centering
\includegraphics[width=0.35\textwidth]{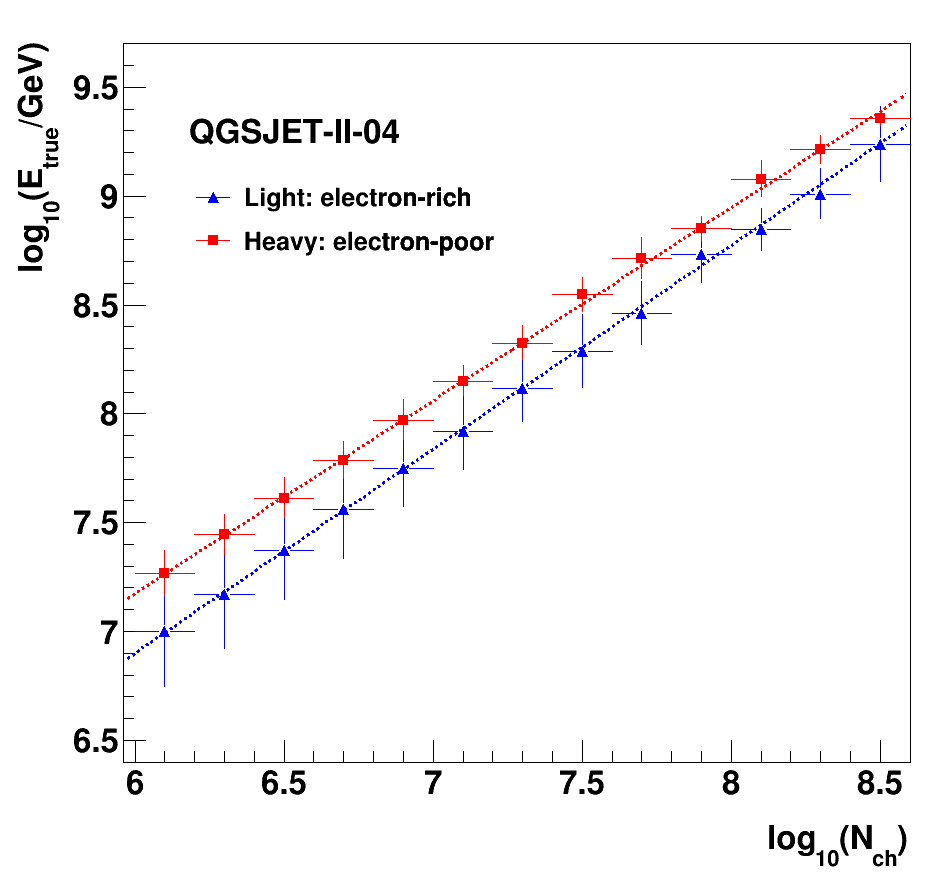}
\hspace{1.5cm}
\includegraphics[width=0.35\textwidth]{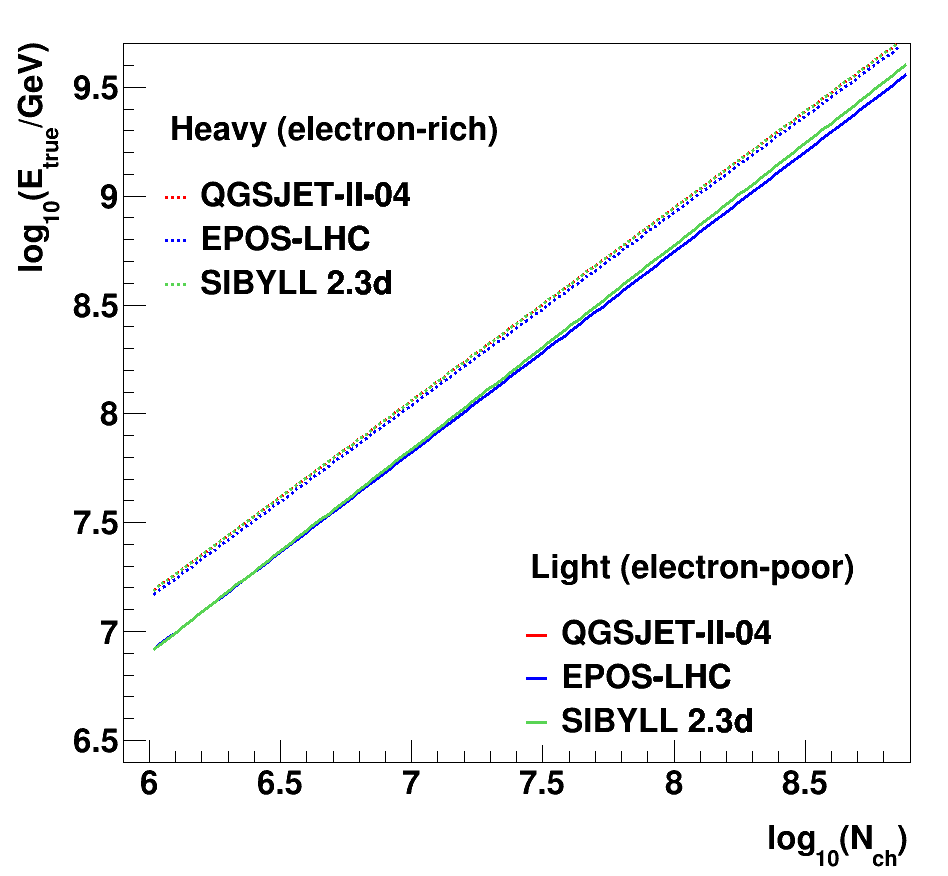}
\caption{Left: The primary energy as a function of the number of charged particles for the hadronic interaction model QGSJET-II-04, along with the linear fit (lines). Right: A comparison of the energy calibration function of heavy (dashed lines) and light (solid lines) primaries for QGSJET-II-04, EPOS-LHC and SIBYLL 2.3d.}
\end{figure}

\subsection{Spectra of heavy and light mass groups}
The energy spectra of heavy and light mass groups are reconstructed by means of the relation $E(N_{ch})$ for two separated samples. I.e. we converted the attenuation corrected shower size into the reconstructed energy, using the model-dependent energy calibration function (Table~1).
Figure 3 shows the resulting reconstructed energy spectra for light and heavy initiated showers with statistical and systematic errors. The all-particle spectrum of primary cosmic rays is obtained by summing the light and heavy components.
Systematic uncertainties in flux are discussed in the next section. The bin-to-bin migration effect is not taken into account yet, but the effect is expected to be small.

\begin{figure}[b]
\centering
\includegraphics[width=0.329\textwidth]{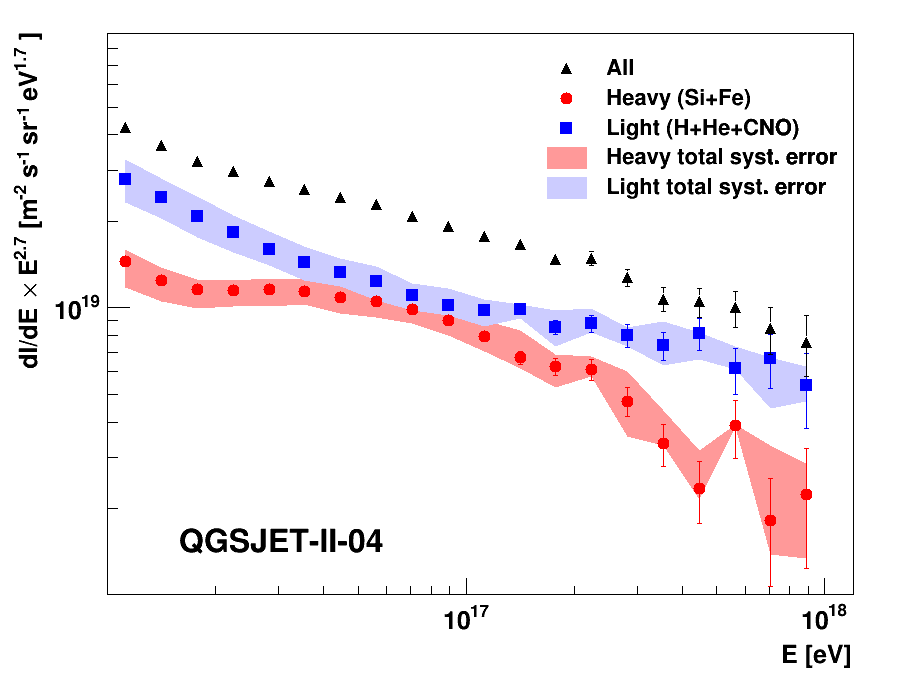}
\includegraphics[width=0.329\textwidth]{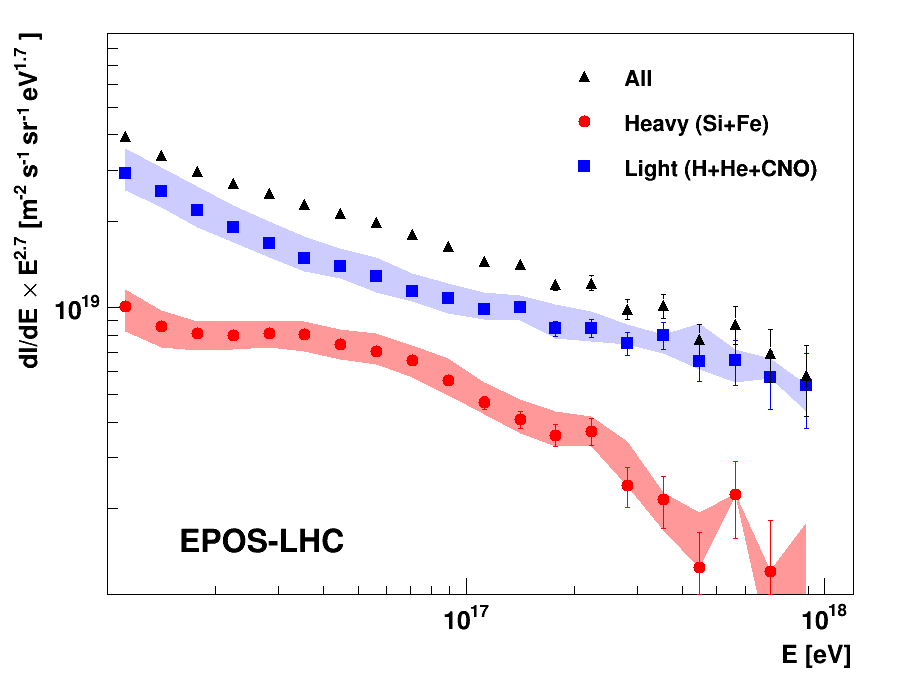}
\includegraphics[width=0.329\textwidth]{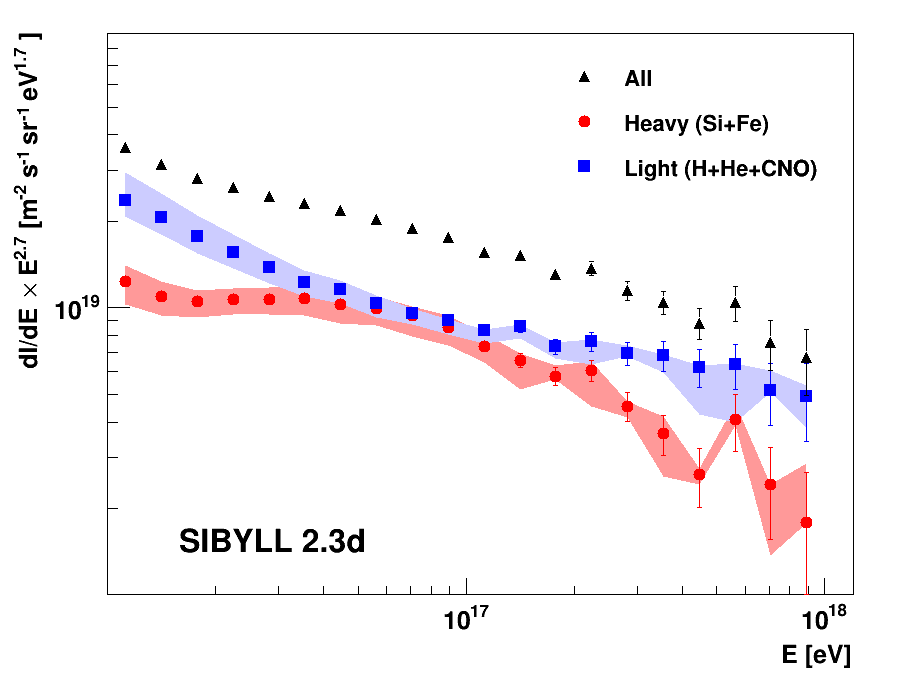}
\caption{
  The resulting primary energy spectra for heavy (Si+Fe) and light (H+He+CNO) primaries based on QGSJET-II-04 (left), EPOS-LHC (middle), and SIBYLL 2.3d (right). All means the all-particle spectrum of primary cosmic rays, which is the sum of the light and heavy mass components.
}
\end{figure}

A broken power-law fitting has been performed for the heavy spectra (Fig.~4). Table~2 summarizes the positions of the spectral breaks, and the slopes before and after the heavy knee. All features observed by the previous analysis are well confirmed.
The spectrum of heavy primaries, i.e.~electron-poor (Si+Fe) events, shows a clear knee-like structure at around $10^{16.7}$~eV for all three models, where the power spectral index changes from $\gamma\sim2.7$ to $\gamma\sim3.3$ for all models.
A remarkable feature is that the concave structure at about $10^{16}$~eV is significantly visible in the spectrum of electron-poor components. In the energy spectrum of the light primaries, a hardening feature above about $10^{16.5}$~eV is observed and the spectral slope changes smoothly.

The comparison of the reconstructed energy spectra of the electron-poor and electron-rich mass groups to other post-LHC models of
QGSJET-II-04, EPOS-LHC, and SIBYLL 2.3d is shown in Fig.~4, where all spectra were reconstructed by applying the CIC technique.
The energy spectra of QGSJET-II-04 and SIBYLL 2.3d show a very similar tendency, while the total flux of EPOS-LHC is shifted by about 10\% due to the different ratio of $N_{ch}/N_{\mu}$. In particular, the EPOS-LHC model predicts more muons than the other post-LHC models, so that the data interpretation leads to a more light composition.
In general, the electron-rich sample is always more abundant due to the separation around the CNO mass group. The muon content might also affect some difference of absolute abundances. However, all the spectra turn out a similar feature of the energy spectrum.
In Fig.~4 (right), the all-particle energy spectrum, which is obtained from the sum of the individual heavy and light spectra, is displayed. The ratio of EPOS-LHC and SIBYLL 2.3d spectra with respect to the QGSJET-II-04 model are shown as well. It presents a small difference among the post-LHC hadronic interaction models over the whole energy range. 

\begin{figure}[t]
\centering
\includegraphics[width=0.455\textwidth]{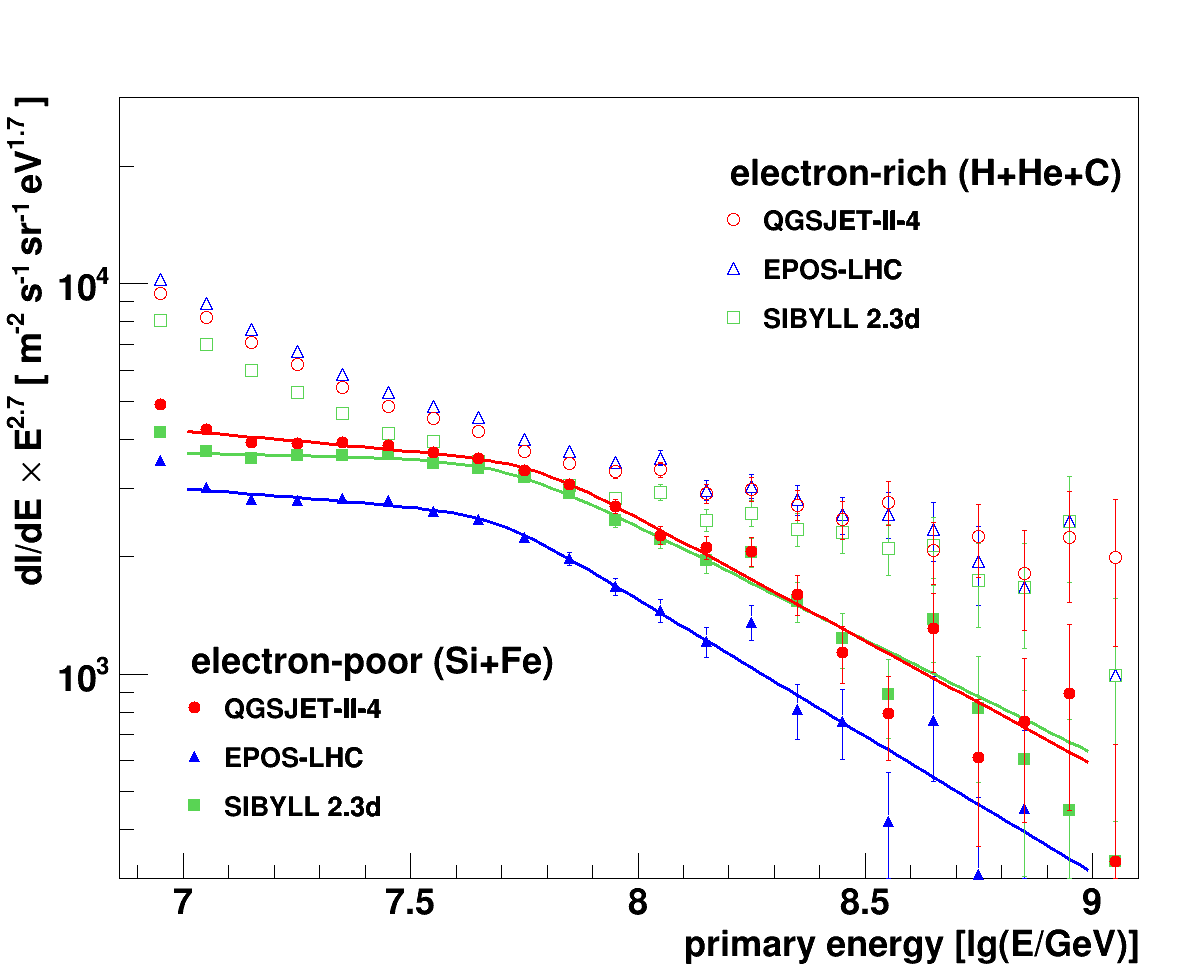}
\hspace{1.3cm}
\includegraphics[width=0.37\textwidth]{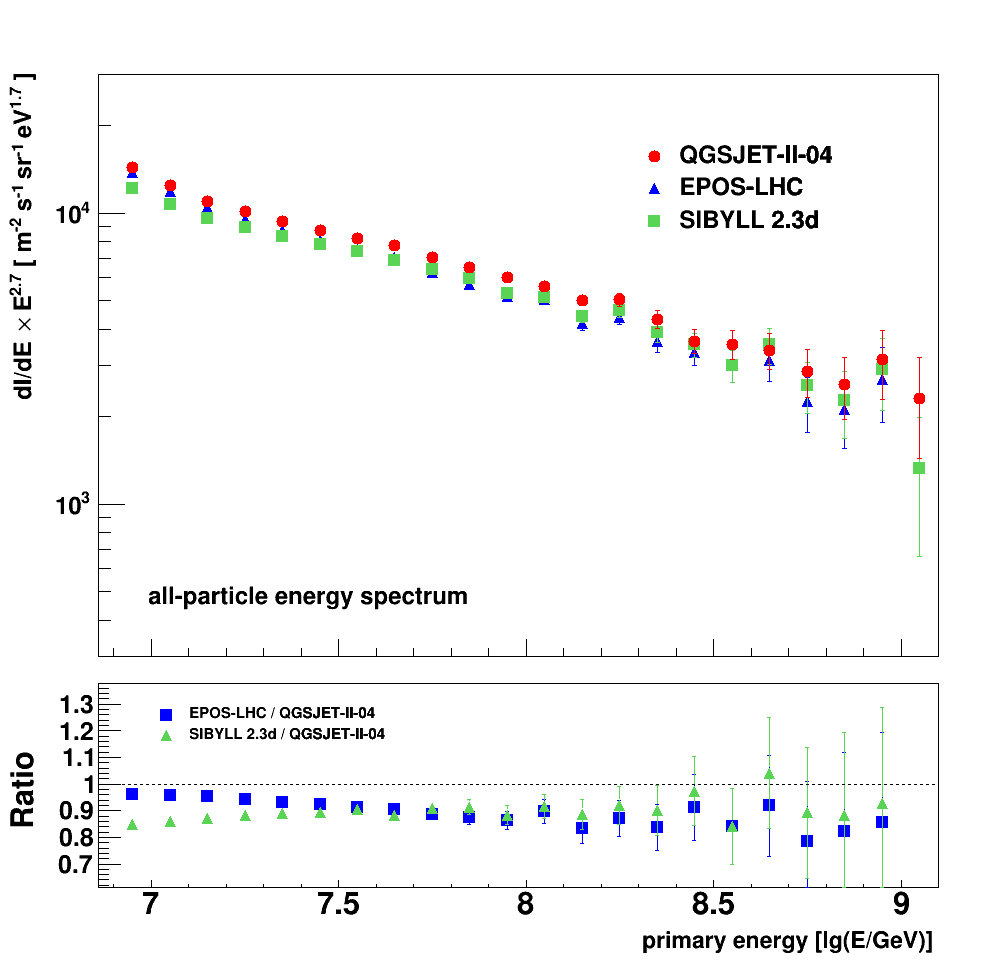}
\caption{
  Left: A comparison of the interaction models QGSJET-II-04, EPOS-LHC, and SIBYLL 2.3d for the reconstructed energy spectra of heavy (Si+Fe) and light (H+He+CNO) primaries. Right: The all-particle energy spectrum based on the three models (top) and the ratio with respect to the QGSJET-II-04 model (bottom). 
}
\end{figure}

\begin{table}[b!]
\begin{center}
\small    
\begin{tabular}{lccccc}
\hline
electron-poor & log$_{10}(E_{k}/{\rm GeV})$ & $\gamma_{1}$ & $\gamma_{2}$ & $\Delta \gamma$ & $\chi^{2}$/ndf \\ \hline
QGSjet-II-04  & 7.78 $\pm$ 0.05  & 2.80 $\pm$ 0.01 & 3.33 $\pm$ 0.06 & 0.53 & 3.92 \\
EPOS-LHC      & 7.71 $\pm$ 0.04  & 2.73 $\pm$ 0.01 & 3.28 $\pm$ 0.06 & 0.55 & 1.96 \\
SIBYLL 2.3d   & 7.69 $\pm$ 0.04  & 2.79 $\pm$ 0.01 & 3.40 $\pm$ 0.06 & 0.61 & 2.70 \\ \hline
\end{tabular}
\caption{The positions of the spectral breaks and the spectral slopes after applying a broken power law fit to the spectra of electron-poor (heavy) events.}
\end{center}
\end{table}

\begin{table}[b!]
  \begin{center}
  \small     
\begin{tabular}{lcccccc}\hline
& \multicolumn{2}{c}{QGSJET-II-04} & \multicolumn{2}{c}{EPOS-LHC} & \multicolumn{2}{c}{SIBYLL 2.3d}\\
&Light(\%) &Heavy(\%) &Light(\%) &Heavy(\%) &Light(\%) &Heavy(\%)\\ \hline
Spectral index: &&&&&&\\
$\gamma$ = -2.8 &2.66 &1.78 &2.54 &3.21 &3.96 &0.63\\
$\gamma$ = -3.2 &5.32 &9.32 &0.28 &2.67 &5.41 &7.13\\ \hline
CIC method: &&&&&&\\
Attenuation fit &$<$0.1 &$<$0.1 &$<$0.1 &$<$0.1 &$<$0.1 &$<$0.1\\ \hline
$\theta_{ref}=10^{\circ}$ &9.80 &7.03 &9.29 &7.22 &9.91 &6.34\\ 
$\theta_{ref}=30^{\circ}$ &14.1 &10.0 &17.5 &14.2 &11.4 &0.48\\ \hline
Energy conversion &0.79 &0.51 &0.58 &0.51 &0.23 &1.35\\ \hline
Total &$^{+14.1}_{-11.5}$ &$^{+10.1}_{-11.8}$ &$^{+17.7}_{-9.32}$ &$^{+14.5}_{-7.71}$ &$^{+9.89}_{-13.9}$ &$^{+1.43}_{-9.66}$\\\hline
\end{tabular}
\caption{\label{label} Systematic uncertainties in flux at $10^{17}$eV.
  The total systematic uncertainty is the sum in quadrature of all terms.}
\end{center} 
\end{table}

\section{Systematic uncertainties}
The possible sources of systematic uncertainties on the reconstructed energy spectrum are uncertainties introduced by the analysis method and procedure, e.g. each fit parameter has an associate error and gives an influence on the energy spectrum.
Some uncertainties could be introduced by the nature of the cosmic rays and the extensive air showers, such as the shower fluctuations. In addition, the reconstruction accuracy of the arrival direction could give systematic uncertainties to the estimation of the energy spectrum.

In the application of the CIC method, a global fit of the attenuation curves are performed, in order to correct the zenith angle dependence of the number of charged particles. Each fit parameter has an associated error and it results in uncertainties in the determination of the number of charged particles for the reference angle of 20$^{\circ}$.
Since the fit parameters are correlated with each other, the propagation errors are calculated to estimate the systematic uncertainty induced by the attenuation curve fit. Systematic errors induced by the attenuation curve fit for the corrected $log_{10}N_{ch}^{20^{\circ}}$ are estimated to be less than 1\% in the full energy range for heavy and light primaries.

The conversion relation is determined with a fit on the distribution of the true primary energy as a function of the number of charged particles (see Fig.~2), which have uncertainties due to the fluctuation of number of charged particles in the MC simulations. Uncertainties from the fit parameters are used to determine the systematic effect on the reconstructed energy spectrum. In the same way of uncertainties of the attenuation curves, the propagation error of the energy conversion function is used.

The shower fluctuations are another source of systematic uncertainties, which is basically caused by the nature of the development of extensive air showers. The influence of these fluctuations on the primary energy spectrum is estimated by using simulation data. The primary energy spectrum of the form $E^{\gamma}$ follows the power law with the spectral index of $\gamma$ = -3, whereas the true primary energy spectrum is characterized by the spectral index of $\gamma$ = -2 in the Monte Carlo simulation data. Therefore, the influences of changing the spectral index of the true primary energy spectrum from $\gamma$ = -2 to $\gamma$ = -3 are estimated here.
In order to obtain the distribution of a true primary energy spectrum with a power law of $E^{-3}$, the original simulation data were weighted with a slope of $\gamma$ = -3. By varying the spectral index of $\pm 0.2$, i.e. changing the effect of fluctuations on the reconstructed spectrum, the systematic uncertainty is estimated. The systematic deviation due to the shower fluctuation is
evaluated by this procedure, e.g. for QGSJET-II-04, to be about 5\% and 9\% for light and heavy primaries, respectively.
For the energy calibration based on the simulation, we selected events only for the zenith angle range of around the reference angle. The reason is that the peak, which was chosen as the reference angle, at the zenith angle distribution has large statistics, so that the systematic effects can be reduced. To estimate the zenith angle uncertainties, simulation data were selected around two different reference angles of 10$^{\circ}$ and 30$^{\circ}$. The differences with a comparison of energy spectrum of $\theta_{ref}$ = 20$^{\circ}$ are included into the total systematic uncertainty.

All these individual systematic contributions were considered to be uncorrelated, and combined thus in quadrature to obtain the total systematic uncertainty (see Table~3). The systematic uncertainty (i.e. sum in quadrature of all terms discussed above except the energy resolution) in the flux is of the order of about $^{+14.1}_{-11.5}$\% for light and $^{+10.1}_{-11.8}$\%for heavy primaries
at the primary energy of $10^{17}$~eV.

\begin{figure}[t!]
\centering
\includegraphics[width=0.6\textwidth]{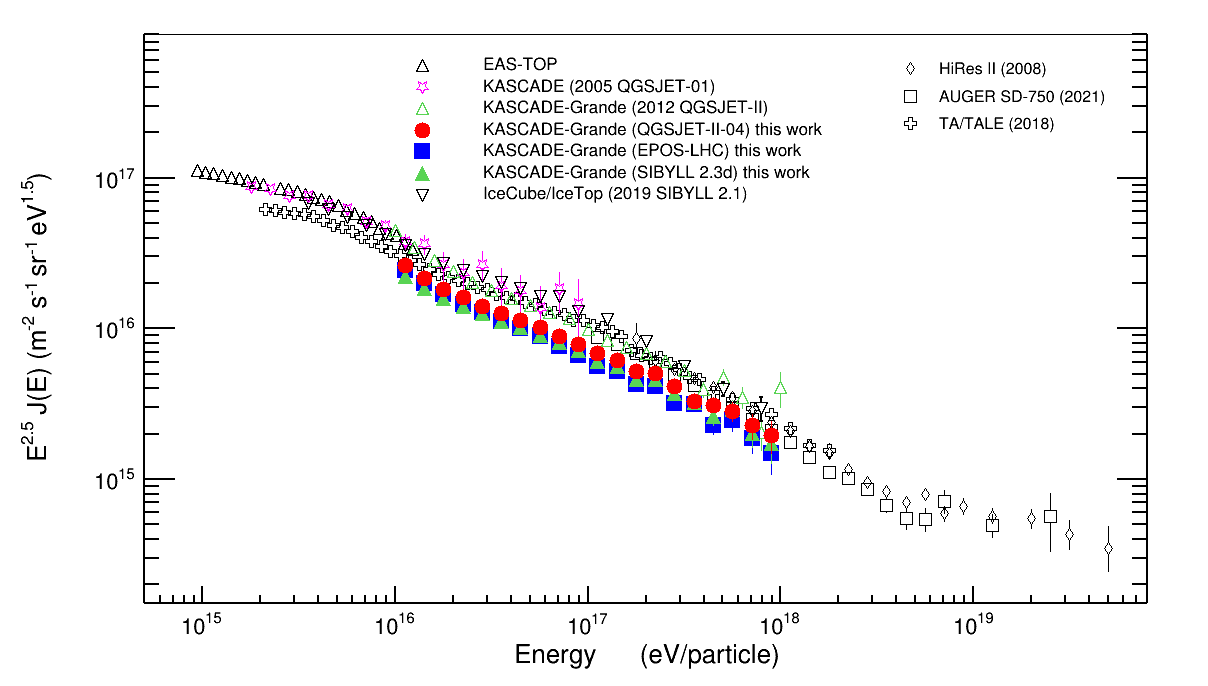}
\caption{
  A comparison of the reconstructed all-particle energy spectrum
  ESA-TOP, IceCube/IceTop, Pierre Auger Observatory, Telescope Array.
  The markers of red circles, blue squares and green triangles
  are the spectra of QGSJET-II-04, EPOS-LHC and SIBYLL 2.3d, respectively.
}
\end{figure}

A comparison of the resulting all-particle energy spectra based on the different post-LHC interaction models with other experiments ESA-TOP, IceCube/IceTop, Pierre Auger Observatory (PAO) and Telescope Array (TA) is presented in Fig.~5. As shown in Fig.~4, the KASCADE-Grande results show relatively small dependencies among the hadronic interaction models in use.
Though, in terms of the absolute flux, there is for KASCADE-Grande an up to 20\% lower flux compared to the other measurements. This might be due to the measurements close to see level.
Moreover, as KASCADE-Grande measures the total number of electrons and muons, separately, these differences are related to the absolute normalization of the energy scale by the various models.
All experiments operates at different observation levels, use different analysis techniques and different hadronic interaction models to interpret their data, nevertheless, a good agreement between the results of the KASCADE-Grande experiment and others is shown in the energy range of PeV to EeV. At the higher energy ($\sim10^{18}$~eV), the KASCADE-Grande result is statistically in agreement with the result of Pierre Auger Observatory.

\section{Muon content studies} 
The muon number of cosmic-ray air showers measured by the KASCADE-Grande experiment is investigated in the primary energy between $10^{16}$ and $10^{18}$~eV for three zenith angle ranges. The hadronic interaction models QGSJET-II-04, EPOS-LHC and SIBYLL 2.3d are used. For the energy calibration, the observed muon-number distributions with respect to Monte Carlo predictions from the data-driven model Global Spline Fit (GSF) and the Pierre Auger energy scale are compared.
The estimation of the total number of muons as a function of the primary energy at ground level with KASCADE-Grande has shown that the muon content for vertical showers is likely to be overestimated by the post-LHC models QGSJET-II-04, EPOS-LHC and SIBYLL 2.3d. However, for more inclined events up to 40$^{\circ}$, the data shows a reasonable agreement with the MC expectations. Further details on the analysis method and the result can be found in Ref. \cite{JuanCarlos}.

\section{Conclusion}
By means of the shower size measured by KASCADE-Grande and the $Y_{CIC}$ technique, the energy spectra of heavy and light mass groups were reconstructed based on the post-LHC hadronic interaction models QGSJET-II-04, EPOS-LHC, and SIBYLL 2.3d.
All structures of the energy spectra confirmed by previous measurements are shown: observation of a heavy knee at around $10^{17}$ eV accompanied by a flattening of the light component at the same energy. This might be a sign of an extra-galactic component and it is already dominant below the energy of $10^{17}$~eV for all post-LHC models.
Lastly, KASCADE Cosmic-ray Data Centre (KCDC) \cite{KCDC} is a web-based platform (https://kcdc.iap.kit.edu), in which scientific data of the completed KASCADE and KASCADE-Grande experiments are made available for the astroparticle community and interested general public. A detailed discussion on the latest release of KCDC can be found in Ref.~\cite{AH_KCDC}.


\begin{thebibliography}{99}
\footnotesize{
\bibitem{KASCADE}
  T. Antoni et al., KASCADE Collaboration,
  Nucl. Instr. Meth. A {\bf 513} (2003) 490
\bibitem{KASCADE-Grande}
  W.D. Apel et al., KASCADE-Grande Collaboration,
  Nucl. Instr. and Meth. A {\bf 620} (2010) 202
\bibitem{KA_APP}
  T. Antoni et al., KASCADE Collaboration,
  Astropart. Phys. {\bf 24} (2005) 1-25
\bibitem{KG_APP}
  W.D. Apel et al., KASCADE-Grande Collaboration,
  Astrop. Phys. {\bf 36} (2012) 183
\bibitem{KG_PRL}
  W.D. Apel et al., KASCADE-Grande Collaboration,
  Phys. Rev. Lett. {\bf 107} (2011) 171104
\bibitem{KG_PRD}
  W.D. Apel et al., KASCADE-Grande Collaboration,
  Phys. Rev. D {\bf 87} (2013) 081101
\bibitem{JuanCarlos}
  J.C. Arteaga-Vel\'azquez et al., KASCADE-Grande Coll., PoS(ICRC2023), these proceedings
\bibitem{CORSIKA}
  D. Heck et al.,
  Report Forschungszentrum Karlsruhe, FZKA 6019 (1998)  
\bibitem{QGS04}
  S. Ostapchenko,
  Phys. Rev. {\bf D83} (2011) 014018
\bibitem{EPOS}
  T. Pierog et al.,
  Phys. Rev. C {\bf 92} (2015) 034906  
\bibitem{SIB23d}
  F. Riehn et al.,
  Phys. Rev. D {\bf 102} (2020) 063002
\bibitem{KCDC}
  A. Haungs, D. Kang et al., KASCADE-Grande Collaboration,
  Eur. Phys. J. C (2018) 78  
\bibitem{AH_KCDC}
  A. Haungs et al., PoS(ICRC2023)1614, these proceedings
}
\end{thebibliography}
\end{document}